\documentclass[prb,superscriptaddress,twocolumn]{revtex4-1}
\usepackage{times,amsmath}
\usepackage{epsfig}
\usepackage{color}
\usepackage{longtable}
\usepackage{graphicx}
\usepackage{dcolumn}
\usepackage{bm}
\usepackage{tabularx}
\usepackage{hyperref}
\usepackage{multirow}
\hypersetup{colorlinks=true, citecolor=blue, filecolor=blue, linkcolor=blue, urlcolor=blue}

\newcommand*{\citen}[1]{%
  \begingroup
    \romannumeral-`\x 
    \setcitestyle{numbers}%
    \cite{#1}%
  \endgroup   
}
\begin{document}
\title{Ternary Inorganic Electrides with mixed bonding}
\author{Junjie Wang}
\affiliation{Materials Research Center for Element Strategy, Tokyo Inst. of Technology, 4259-SE3 Nagatsuta-cho, Midori-ku, Yokohama, Kanagawa, 226-8501, Japan.}
\affiliation{International Center for Materials Discovery, State Key Laboratory of Solidification Processing, Northwestern Polytechnical University, Xi'an, Shanxi 710072, People's Republic of China}

\author{Qiang Zhu}
\email{qiang.zhu@unlv.edu}
\affiliation{Department of Physics and Astronomy, High Pressure Science and Engineering Center, University of Nevada, Las Vegas, NV 89154, USA}

\author{Zhenhai Wang}
\affiliation{Skolkovo Institute of Science and Technology, Skolkovo Innovation Center, 3 Nobel St., Moscow 143026, Russia}

\author{Hideo Hosono}
\affiliation{Materials Research Center for Element Strategy, Tokyo Inst. of Technology, 4259-SE3 Nagatsuta-cho, Midori-ku, Yokohama, Kanagawa, 226-8501, Japan.}

\date{\today}
\begin{abstract}
A high-throughput screening based on first-principles calculations was performed to search for new ternary inorganic electrides. From the available materials database, we identified three new thermodynamically stable materials (Li$_{12}$Mg$_3$Si$_4$, NaBa$_2$O and Ca$_5$Ga$_2$N$_4$) as potential electrides made by main group elements, in addition to the well known mayenite based electride (C12A7:$e^{-}$). Different from those conventional inorganic electrides in which the excess electrons play only the role of anions, the three new materials, resembling the electrides found in simple metals under high pressure, possess mixed ionic and metallic bonding. The interplay between two competing mechanisms, together with the different crystal packing motifs, gives rise to a variety of geometries in anionic electrons, and rich physical phenomena such as ferromagnetism, superconductivity and metal-insulator transition. Our finding here bridges the gap between electrides found at ambient and high pressure conditions.

\end{abstract}
\maketitle

\section{Introduction}
Electrides are said to be a class of unconventional compounds which contain excess valence electrons confined in the void or interlayer space and playing the role of anions \cite{Dye-ACR-2009}. The first stable electride was achieved by Matsuishi et al by removing two oxygen ions from the cages of precursor compound, Ca$_6$Al$_7$O$_{16}$ (C12A7) \cite{Matsuishi-Science-2003}. Due to its room temperature stability, C12A7:$e^{-}$ enables many technological applications such as the splitting of carbon dioxide at room temperature \cite{Toda-NC-2013}, synthesis of ammonia from atmospheric nitrogen under mild conditions \cite{Kitano-NChem-2012} and using as a low electron-injection barrier for organic light-emitting diodes (OLEDs)\cite{Hosono233}. Recently, a number of new electrides have been increasingly discovered in experiment including 2D (Ca$_2$N \cite{Lee-Nature-2013}, Y$_2$C\cite{Zhang-CM-Y2C-2104}), 1D (Y$_5$Si$_3$ \cite{Lu-JACS-2016}, Sr$_5$P$_3$\cite{Wang-JACS-2017}) and 0D (YH$_2$ \cite{mizoguchi2016hydride}). Depending on the connectivity of crystal cavities and channels, the identified electrides could be classified to zero, one and two dimensions (0D, 1D and 2D). Due to the loosely bounding nature, the confined electrons form (partially occupies) bands close to Fermi level, which could lead to dramatically reduced work function and a variety of electronic properties. It is worthy mention that recent researches reported that some electrides can be good platforms for toplogical materials \cite{Hirayama-PRX-2018, Huang-NanoL-2018, Ca3Pb-2018}. Thanks to the growing supercomputing power and advances in computational methodology, computation driven materials design began to play the increasingly important role in the discovery of new electride materials. A number of candidate materials have been proposed recently either by computational screening over the materials database \cite{Inoshita-PRX-2014, Tada-IC-2014}, or more advanced first principles crystal structure prediction approaches \cite{Wang-JACS-2017, Ming-JACS-2016, Zhang-PRX-2017}.  

In the research on inorganic electrides, the conventional wisdom is to create the condition of excess electrons on the parent ionic compounds by removing the anion species (O or H). The resulting compounds with unbalanced charges may be stabilized due to the rearrangement of the excess electrons. Although a number of metal rich materials have been found, only a few of them are electrides. For instance, rubidium and cesium form many stable metal-rich oxides, but only Cs$_3$O was proposed to be an electride recently \cite{Park-PRL-2018}. To make an electride, the crystal structure must provide notable crystal cavities to accommodate the excess electrons. Furthermore, the metal-metal bonding and the confinement effect of the electrons competes with each other when the excess electrons are present. Usually, the chemical bonding of materials with high concentration of metals tends to be governed by metal-metal  bonding. To date,  all the experimentally confirmed electrides so far (C12A7:$e^{-}$, Ca$_2$N \cite{Lee-Nature-2013}, Y$_2$C\cite{Zhang-CM-Y2C-2104}, Y$_5$Si$_3$ \cite{Lu-JACS-2016}, Sr$_5$P$_3$\cite{Wang-JACS-2017}), possess slightly metal rich stoichiometries adjacent to the charge balanced compounds. Our current knowledge on the inorganic electride seems to be limited to a rather narrow stoichiometry range governed by the chemical valence rule.

\begin{figure*}
\includegraphics[width=0.75\linewidth]{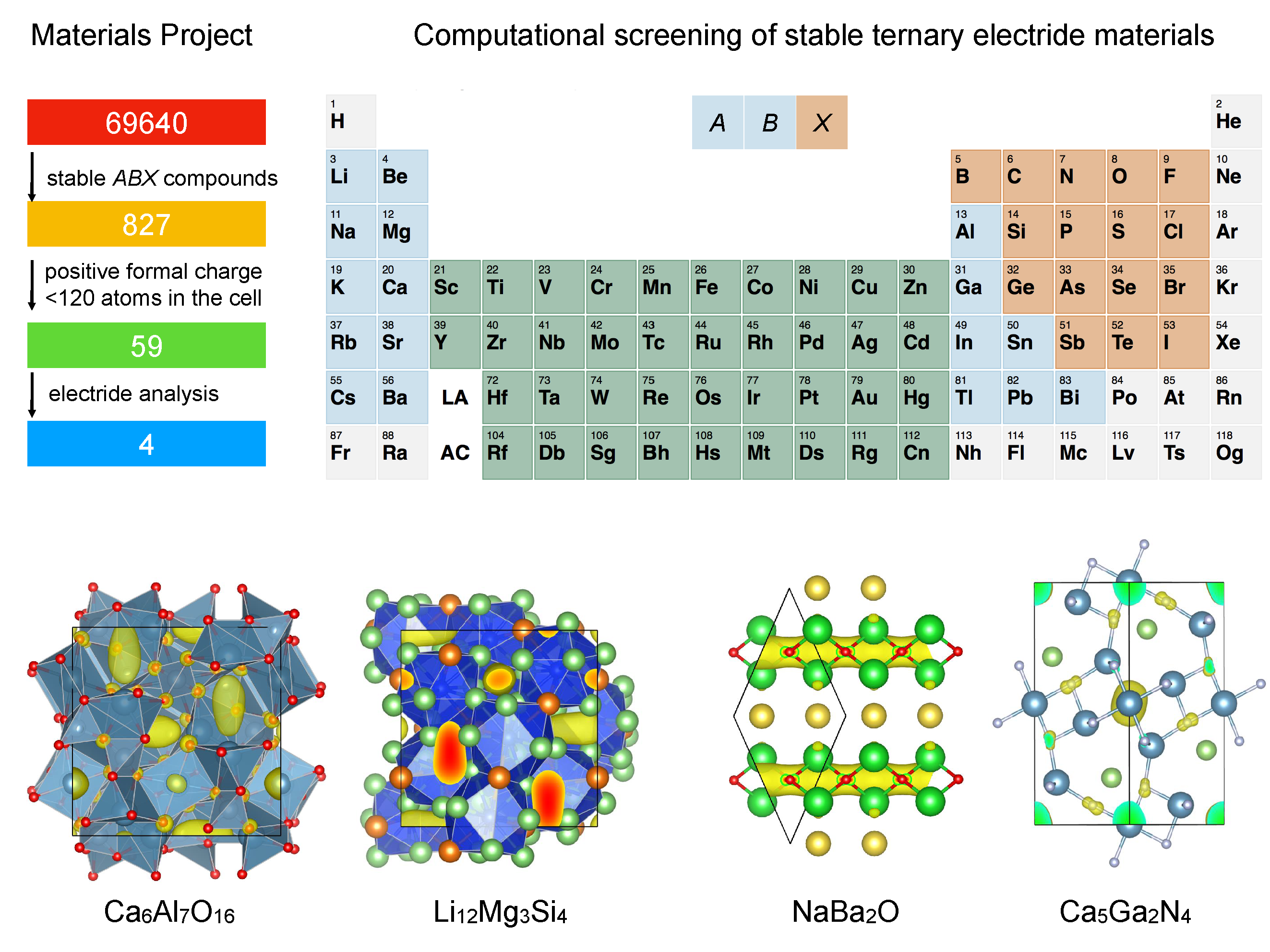}
\caption{\label{fig0} {The proposed scheme of screening electride materiala from the available on the Materials Project and
	the identified four stable ternary electrides 
	with the highlight of isosurfaces of partial charge density in the range of -1.0 $< E-E_{\textrm F} <$ 0 eV in this study.}}
\end{figure*}

However, the recent high pressure studies substantially expand our understanding on electrides. In many simple metals (Ca, Li, Cs, Na, etc) under compression, the extreme case of metal rich materials, the valence electrons no longer follow the nearly free electron approximation, but are forced away from the near core regions to interstitial regions, due to the combined effect of Coulomb repulsion, Pauli exclusion and orthogonality. Rousseau and Ashcroft \cite{Rousseau-PRL-2008} constructed a model to quantitatively describe the tendency of interstitial electronic localization in fcc and bcc crystals based on the ratio $r_c/r_s$, where $r_c$ is the radius of metal core sphere and $r_s$ is the Wigner-Seitz radius. Miao and Hoffman \cite{Miao-ACR-2014} further developed a unified theory by treating interstitial electrons as interstitial quasi atom (ISQ) and filling them quantized orbitals. According to the ISQ model, there exist competition between valence orbitals of atoms and ISQ, and potentially between ISQ and $s$-$d$ switch for heavy alkali and alkaline earth metals (K, Rb, Cs, Ca, Sr, Ba) \cite{Miao-ACR-2014, Hosono-PTRSA-2015}. These high pressure electrides (HPEs) have been found to exhibit a variety of intriguing physical properties, such as metal-semiconductor/insulator transition under pressure \cite{Ma-Nature-2009, Lv-PRL-2011, Pickard-PRL-2009, Matsuoka-Nature-2009}, superconductivity \cite{Shimizu-Nature-2002,Yao-PRB-2009}. Some HPEs were also predicted to possess covalent \cite{Miao-ANIE-2017}, metallic bondings \cite{Miao-JACS-2015} and ferromagnetism \cite{Pickard-PRL-2011}. Despite these appealing properties, all HPEs are not quenchable to the ambient conditions, which limits further experimental study and material utilization.

From the aspect of excess electrons, it is obvious that HPEs, resembling the extremely electron rich materials, are very different compared to the traditional electrides found at ambient conditions. Can we find the stable electrides which resembles the features of HPE and utilize them under ambient condition? A common strategy is to replace high pressure with chemical pressure. The materials found at high pressure could be materialized  by \textit{chemical pre-compression} at ambient condition. 

Considering the binary compounds, we are not aware of such materials to our knowledge, though they have been thoroughly investigated recently \cite{Tada-IC-2014, Zhang-PRX-2017, Ming-JACS-2016}. This motivates us to extend the search to materials with ternary systems. There have been a few computational efforts attempting to find ternary electrides so far, but only a few metastable hypothetical compounds were mentioned in the literature \cite{Tada-IC-2014, Ming-JACS-2016}. To date, direct first-principles crystal structure for multi-component system in vast chemical space remains a grand challenge at the moment. Hence we choose to start with screening the existing materials from the available database. In the present work, we performed a high-throughput screening on the existing thermodynamically stable ternary compounds made by main group elements with a specified chemical space, from which three new materials (Li$_{12}$Mg$_3$Si$_4$, NaBa$_2$O and Ca$_5$Ga$_2$N$_4$) as potential electrides, in addition to the well known mayenite based electride C12A7:$e^{-}$. These ternary electrides are characterized by a mixed bonding between ionic and metallic. Due to the competition between these two mechanisms, these materials exhibit a diverse range of crystal packings and distributions of anionic electrons, which thus give rise to a variety of physical phenomena such as ferromagnetism, superconductivity and metal-insulator transition. The close analogy between these materials and HPEs will be discussed as well in the following sections.

\section{Computational Methods}

\begin{figure*}
\includegraphics[width=0.80\linewidth]{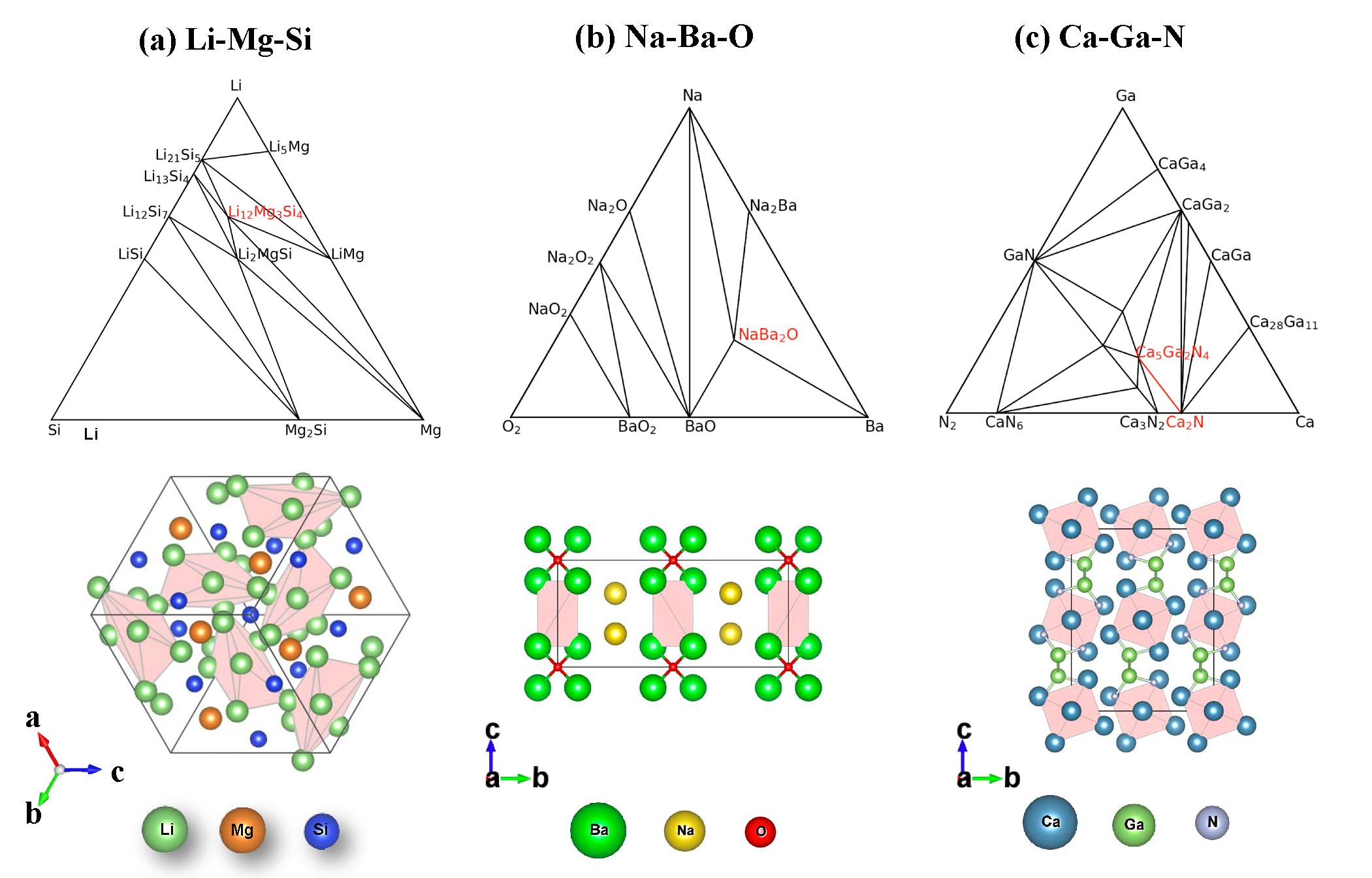}
\caption{\label{fig1} {Three proposed electride materials in the present study. 
        (a) Li-Mg-Si phase diagram (upper panel) and the crystal structure of ternary compound Li$_{12}$Mg$_3$Si$_4$ (lower panel)
	(b) Na-Ba-O phase diagram (upper panel) and the crystal structure of ternary compound NaBa$_2$O (lower panel).
	(c) Ca-Ga-N phase diagram (upper panel) and the crystal structure of ternary compound Ca$_5$Ga$_2$N$_4$ (lower panel).
	In the upper plots, all identified electrides are highlighted by red color in the upper plots.
	In the lower plots, the notable crystal voids are highlighted by red polyhedra.}
	}
\end{figure*}

In the present study, we targeted at all existing thermodynamically stable ternary compounds $A$-$B$-$X$ made of main group elements up to bismuth which are registered in Materials Project database (\url{https://materialsproject.org}) \cite{MP-APL-2013} (see Fig. \ref{fig0}).
This reduced the approximately 69640 structural entries to 827.
These materials were further filtered by the following constraints: 
1) formal charge by the given chemical formula unit needs to be positive, 
2) the maximum number of atoms in the primitive unit cell is less than 120, 
from which 59 candidate structures remain under consideration (see the complete list in the supplementary materials \cite{SI}). We developed an automated scheme to perform the calculation and geometrical analysis on the partial charge density in the range of -0.5 $< E-E_{\textrm F} <$ 0 eV. If a material contains a significant charge density basin ($\geq$ 0.1 $e$ base of Bader partition \cite{Bader-1990}) which does not belong to either nuclear or shared electrons by covalent bonds, it will be considered as the candidate of electride. Followed by a detailed electronic structure analysis to exclude the case where the metallic bonding plays the major role, we successfully identified four materials which are likely to be true electrides, which are C12A7:$e^{-}$, Li$_{12}$Mg$_3$Si$_4$, NaBa$_2$O and Ca$_5$Ga$_2$N$_4$. All of these materials have been experimentally synthesized and represent a variety of crystal packing and chemical bonding, suggesting that plenty of new electrides may have been overlooked in the past.

\begin{table}
	\caption{The crystallographic data of the interstitial electrons for all investigated systems 
	and the corresponding Bader charges. The details of Bader charge calculation are given in Sec. \ref{analysis}.}
	\begin{tabular}{ccccc}
\hline
\hline
	System                & Space group       & Wyckoff & Coordinates  & Charge \\
\hline
	C12A7:$e^{-}$  & $I\bar{4}$3$d$    & 12b     & (0.000, 0.250, 0.875) & 0.26 \\
	Li$_{12}$Mg$_3$Si$_4$ & $I\bar{4}$3$d$    & 12a     & (0.000, 0.500, 0.250) & 0.48 \\
	NaBa$_2$O             & $Cmme$            & 4b      & (0.250, 0.000, 0.500) & 0.55 \\
	Ca$_5$Ga$_2$N$_4$     & $Cmca$            & 4a      & (0.000, 0.000, 0.000) & 0.44 \\
\hline
\hline
\end{tabular}
\label{table1}
\end{table}

\begin{table}
	\caption{The comparisons of lattice constants between experiment and simulation.}
	\begin{tabular}{lllc}
\hline
\hline
	System                & Simulation   & Experiment  &Reference \\
	                      & ~~~~~~(\AA)  & ~~~~~~(\AA)   &\\
\hline
	Li$_{12}$Mg$_3$Si$_4$ & a=10.685     & a=10.688 & [\citen{Pavlyuk-1992}]     \\
 &    &      &   \\
	NaBa$_2$O             & a=6.672          & a=6.591     &  [\citen{NaBa2O-2001}]    \\
                          & b=15.567         & b=15.327 &     \\
                          & c=6.978          & c=6.936  &  \\
 &    &      &   \\
 
	Ca$_5$Ga$_2$N$_4$     & a=4.881     &  a=4.873    & [\citen{Ca5Ga2N4-1988}]\\
                          & b=11.197    &  b=11.105   & \\
                          & c=14.327    &  c=14.217   & \\

\hline
\hline
\end{tabular}
\label{table2}
\end{table}

Table \ref{table1} summarizes the crystallographic information for all four materials.
Although these structures are crystallographically different, they are all featured by the presence of large crystal voids, which are potentially capable of capturing exceeding electrons as anions. In order to analyze their structural properties, we carried out calculations in high precision settings using the plane wave density functional theory (DFT) program VASP \cite{Vasp-PRB-1996}. 
The exchange-correlation functional is described by the generalized gradient approximation in the Perdew-Burke-Ernzerhof parameterization (GGA-PBE) \cite{PBE-PRL-1996}. 
We used the cutoff energy of 600 eV and k-point mesh resolution of around 2$\pi$ $\times$ 0.02 \AA$^{-1}$ for all calculations. The lattice constants and atomic coordinates were fully relaxed until the force acting on each ion is less than 0.001 eV/\AA. The comparisons between our simulated and experimental lattice constants are shown in Table \ref{table2}. The excellent agreement encouraged us to use the same parameters to analyze the electronic properties. For the calculation of projected density of states, we used the following Wigner-Seitz radius values: 2.00 \AA ~(Ba), 1.80 \AA ~(Na), 1.80 \AA ~(Ca), 1.70 \AA ~(Ga), 0.80 \AA ~(O), 0.75 \AA~(N).

\begin{figure*}
\epsfig{file=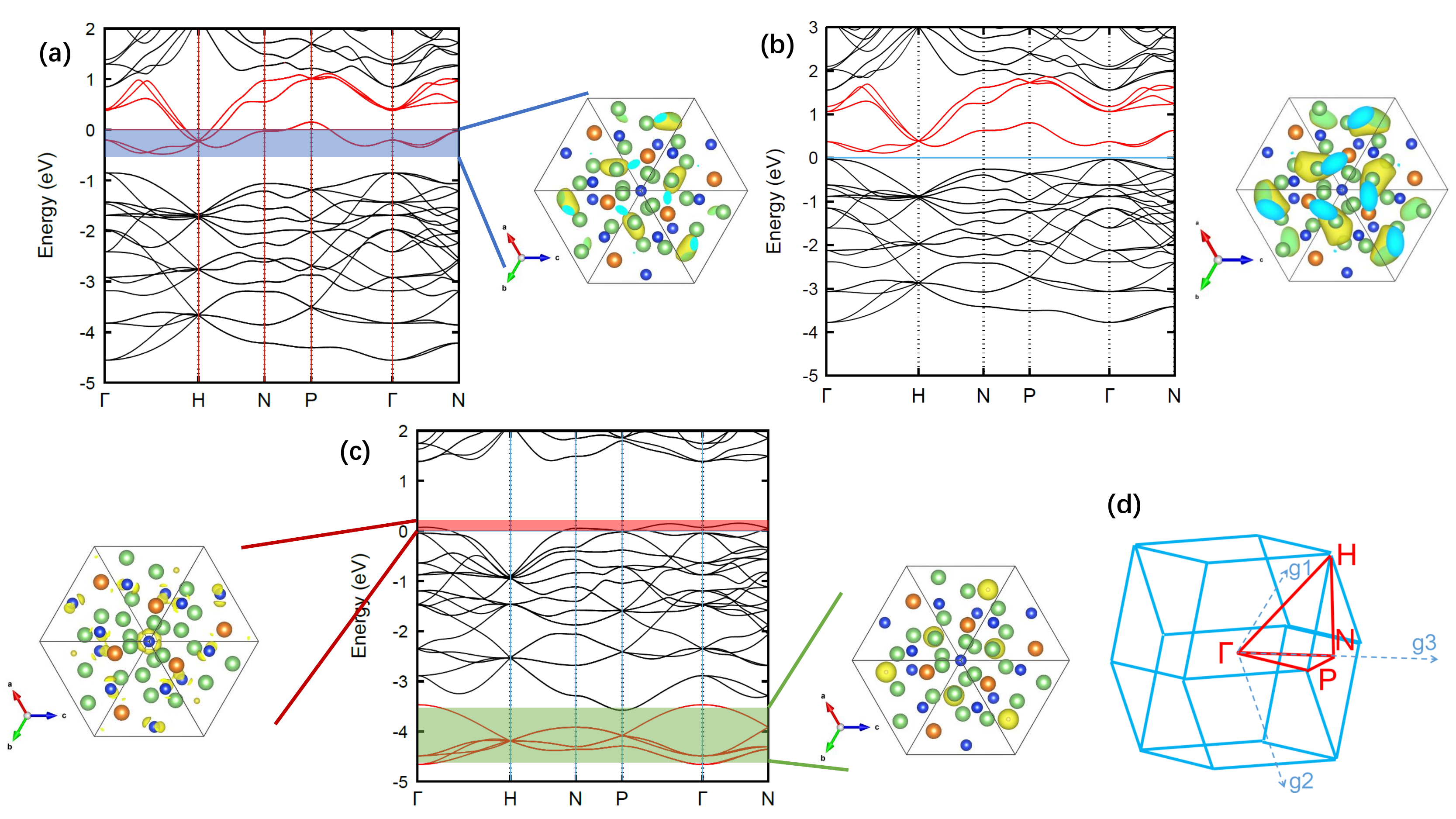, width=0.80\textwidth}
	\caption{\label{limgsi}
	The calculated electronic structures of Li$_{12}$Mg$_3$Si$_4$:$2e^{-}$ and artificial Li$_{12}$Mg$_3$Si$_4$H$_3$. 
	(a) The calculated band structure and partial charge density (-0.5 $< E-E_{\textrm F} <$ 0 eV) of  Li$_{12}$Mg$_3$Si$_4$;
    (b) The calculated band structure and partial charge density empty states at lattice voids (highlighted in red in band structure) of Li$_{12}$Mg$_3$Si$_4$-$4e^{-}$;
	(c) The calculated band structure and partial charge densities in the ranges of -5 eV$< E-E_{\textrm F} <$ -3.5 eV and 0$< E-E_{\textrm F} <$ 0.2 eV of  Li$_{12}$Mg$_3$Si$_4$H$_3$;
    (d) The Brillouin zone of  Li$_{12}$Mg$_3$Si$_4$:$2e^{-}$ and artificial Li$_{12}$Mg$_3$Si$_4$H$_3$. 
    The isosurface values for the partial charge densities are respectively 2e-3 $e$/Bohr$^{3}$ (a), 2e-3 $e$/Bohr$^{3}$ (b), 6e-5 $e$/Bohr$^{3}$ (left panel of (c)), 0.02 $e$/Bohr$^{3}$ (right panel of (c)). The red lines in (d) indicate the K-path used for the band structure plotting in (a) and (b).
	}
\end{figure*}
\section{Results and Discussions}
\subsection{Analogy between Li$_{12}$Mg$_3$Si$_4$:2$e^{-}$ and C12A7:$e^{-}$}
Among these screened structures, Ca$_6$Al$_7$O$_{16}$, known as C12A7:$e^{-}$ in cubic space group ($I\bar{4}3d$, see Table \ref{table1}), is the first known room-temperature and air-stable inorganic electride \cite{Matsuishi-Science-2003}. This compound has also been extensively studied by theorists as well \cite{Li-ANIE-2004, Li-Review-2016, dale-2018}. Ca$_6$Al$_7$O$_{16}$ has a porous structure with 12 crystallographically equivalent notable cages (Wyckoff position 12b) in the unit cell. The excess electrons are thus uniformly distributed among these cages in the cubic crystal structure, which gives 0D electride. Interestingly, Li$_{12}$Mg$_3$Si$_4$ shares the same space group symmetry with C12A7:$e^{-}$ in a denser closed packing manner. Similar to C12A7:$e^{-}$, Li$_{12}$Mg$_3$Si$_4$ has also 12 equivalent cages (12a) (see Table \ref{table1}) in the unit cell (though the porosity is smaller, see Fig.\ref{fig0}). 

C12A7:$e^{-}$ and Li$_{12}$Mg$_3$Si$_4$ has respectively +1 and +2 formal charges for each chemical formula, implying the presence of two and four extra electrons per primitive cell. Therefore we follow the convention to name the latter as Li$_{12}$Mg$_3$Si$_4$:$2e^{-}$. We firstly analyzed the electronic structure of Li$_{12}$Mg$_3$Si$_4$:$2e^{-}$ and checked if it is a true electride. Fig. \ref{limgsi}a shows the calculated band structure of Li$_{12}$Mg$_3$Si$_4$:$2e^{-}$. There exist six partially occupied bands across the Fermi level, which correspond to the six vacant sites per primitive cell of Li$_{12}$Mg$_3$Si$_4$:$2e^{-}$ (right panel of Fig. \ref{limgsi}a). To confirm its electride nature, we artificially constructed a Li$_{12}$Mg$_3$Si$_4$:$2e^{-}$ structure by removing four electrons from the primitive cell of Li$_{12}$Mg$_3$Si$_4$:$2e^{-}$ and calculated its electronic structures (Fig. \ref{limgsi}b). It shows that the six vacant sites became completely empty and a narrow gap appears after the removal of four electrons. Sum of the square of the wave functions corresponding to the empty interstitial band of Li$_{12}$Mg$_3$Si$_4$:$2e^{-}$ is shown in the right panel of Fig. \ref{limgsi}b. One can see that the empty sites are ready to be filled with extra electrons. Furthermore, we introduced six H atoms to the primitive cell of Li$_{12}$Mg$_3$Si$_4$:$2e^{-}$. As expected, very little lattice change happened after relaxation and hydrogen-related bands appear at around -4 eV for the H$^{-}$ (shown in red in Fig. \ref{limgsi}c). However, the Li$_{12}$Mg$_3$Si$_4$H$_3$ turned to be a $p$-type metal with the introduction of H atoms at the interstitial sites. It can be understood that H has a stronger electronegativity than Si to guarantee the formation of singlet H 1$s$ states, i.e., H$^-$ (right panel of Fig. \ref{limgsi}c). Consequently, holes (electron-deficient states) can be formed around Si atoms as shown in the left panel of Fig. \ref{limgsi}c. 

\begin{figure*}[ht]
\epsfig{file=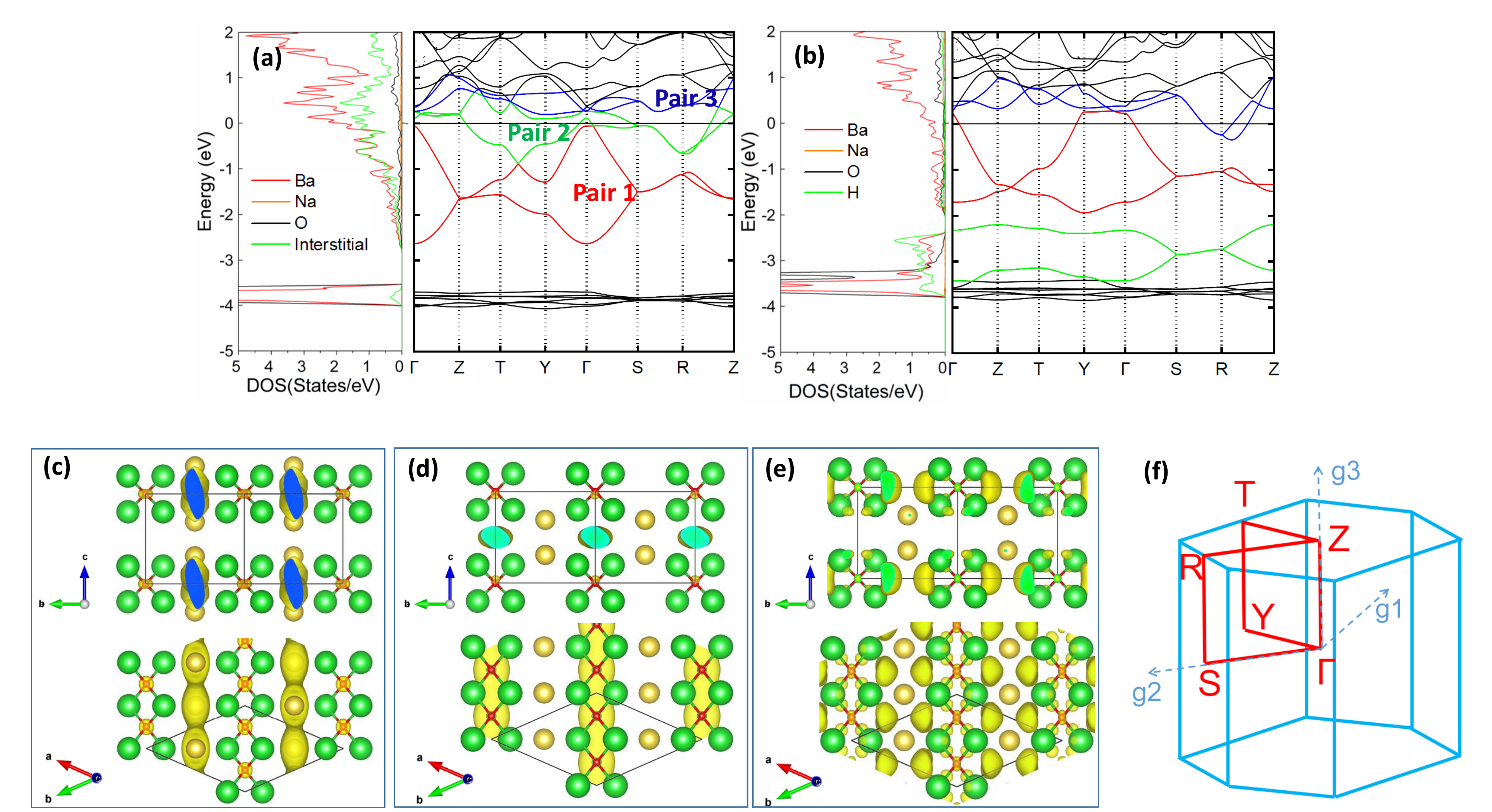, width=0.85\textwidth}
	\caption{\label{nabao}
	The calculated electronic structures of NaBa$_2$O and artificial compound NaBa$_2$OH.
	(a) The band structure and projected density of states of NaBa$_2$O; 
	(b) The band structure and projected density of states of NaBa$_2$OH; 
	and the partial charge densities of the bands (c) Pair 1 (d) Pair 2 and (e) Pair 3 in (a). 
	(f) The Brillouin zone of the NaBa2O and NaBa$_2$OH. The red lines in (f) indicate the K-path used for the band structure plotting in (a) and (b).
	The isosurface value for the partial charge densities in (c), (d) and (e) is 2e-3 $e$/Bohr$^{3}$.}
\end{figure*}

We know that C12A7:$e^{-}$ exhibits metallic conductivity because it only confines 1/3 electrons per cage site. Here the four extra electrons of one Li$_{12}$Mg$_3$Si$_4$:$2e^{-}$ primitive cell can be shared by the six vacant sites. Therefore, Li$_{12}$Mg$_3$Si$_4$:$2e^{-}$ can serve as another model structure to check the conductivity with 2/3 electrons occupancy per site. In addition to C12A7:$e^{-}$ and Li$_{12}$Mg$_3$Si$_4$:$2e^{-}$, another electride phase $cI$16-Li, a high pressure form of elemental Li, possess the same space group $I\bar{4}$3$d$. The close structural similarity between C12A7:$e^{-}$ and $cI$16-Li has been discussed into details in Ref. \cite{Hosono-PTRSA-2015}. It was found that these two materials also shares similar electronic band structure due to structural similarity. C12A7:$e^{-}$ is the first superconducting electride phase found at the ambient conditions \cite{miyakawa2007superconductivity}. However, the critical superconducting transition temperature $T_\textrm{c}$ of C12A7:$e^{-}$ ($\sim$ 0.2 K) is much lower than that of $cI$16-Li ($\sim$20 K). This was attributed to the low DOS at Fermi level. It is expected that the pressurization of both C12A7:$e^{-}$ and Li$_{12}$Mg$_3$Si$_4$:$2e^{-}$ can also enhance their $T_\textrm{c}$ values.

\subsection{Suboxide electride NaBa$_2$O}
The crystal structure of NaBa$_2$O has an orthorhombic unit cell with space group $Cmme$ (see Table \ref{table1}).  In general, the structure can be described by the distorted body-centered cubic (bcc) packing of Na and Ba atoms, in which every third layer is occupied by Na. In the distorted bcc structure, half of the [Ba$_4$] tetrahedral interstitials are filled by O, while the other half interstitials are empty (see Fig. \ref{nabao}). The most striking feature in NaBa$_2$O is that Ba is tetrahedrally coordinated to O instead of the more common [BaO$_6$] octahedra unit observed in other materials. As shown in Fig. \ref{nabao}, the [Ba$_4$O] tetrahedra mutually share the edges, leading to an infinite [Ba$_2$O] chain running through [1$\overline{1}$0] direction, and they are separated by Na atoms along [110] and the empty space along [001]. According to the formal charge state, this compound can be expressed as NaBa$_2$O:$3e^{-}$. Consequently, six extra electrons can be expected in the primitive cell of NaBa$_2$O:$3e^{-}$. In contrast to the abundant interstitial sites in Li$_{12}$Mg$_3$Si$_4$H$_3$:$2e^{-}$ (six sites for four extra electrons, Fig. \ref{limgsi}a), Fig. \ref{nabao} shows that the two interstitial sites in one primitive cell of NaBa$_2$O:$3e^{-}$ need to respectively accommodate six extra electrons.

The calculated electronic band structures and projected densities of states (PDOS) of Ba$_2$NaO are plotted in Fig. \ref{nabao}(a).  The existence of two bands crossing Fermi level suggests its metallic characteristics. To identify the contributions from interstitial electrons, we placed pseudoatoms with a Wigner-Seitz radius of 1.84 \AA ~at the interstitial sites of in the primitive unit cell (see table \ref{table1}) and projected the portions of the wave functions within these spheres to obtain the PDOS curves for the interstitial site shown in Fig. \ref{nabao}a. Indeed, the contributions from the interstitial sites are dominant around Fermi level, which is a typical character of electrides. But this picture is more complicated than other known electrides like Ca$_2$N \cite{Lee-Nature-2013} and Sr$_5$P$_3$ \cite{Wang-JACS-2017}, due to the connection with neighboring bands. To further confirm the electron accumulations around Fermi level, the partial charge densities of the six bands around Fermi level were calculated (Fig. \ref{nabao}c-e). Clearly, bands Pair 1 shows the feature of metallic bonding between the Na atoms. The partial charge density of Pair 2, which cross the Fermi level, shows 1D electride characteristic along the channel surrounded by Ba atoms. And the bands Pair 3 are unoccupied and possess a feature of anti-bonding states of metallic bonding. Another key characteristics of electrides is the negligible lattice strain due to insertion/extraction of anionic ions. Here we inserted two hydrogen atoms at the interstitial sites and relaxed. We found very little lattice change can be observed. Moreover, the calculated band structure of the artificial NaBa$_2$OH shows that the bands Pair 2 disappeared after the introduction of H atoms. Instead, a pair of bands appear at the energy range of around -3.5 to -2.5 eV. It is noteworthy that the degeneracy of the disappeared electride bands is essentially the same as that of the H 1$s$ bands of NaBa$_2$OH.

\subsection{Ferromagnetic electride phase Ca$_5$Ga$_2$N$_4$}
\begin{figure}
\epsfig{file=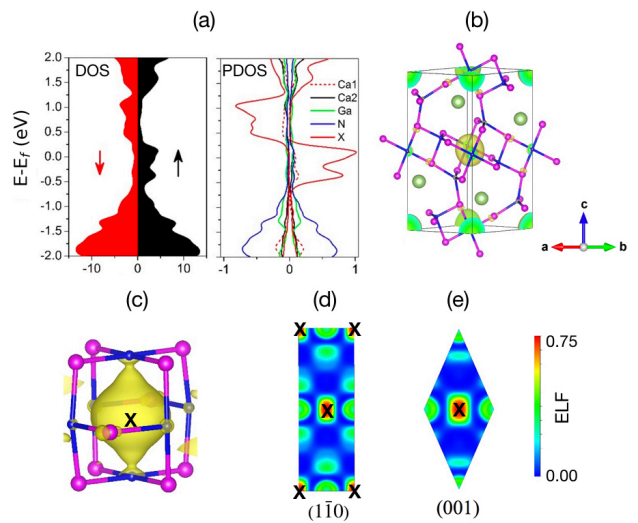, width=0.50\textwidth}
	\caption{\label{fig4}
	The calculated electronic structures of Ca$_5$Ga$_2$N$_4$.
	(a) The total density of states (DOS) and the projected density of states (PDOS) on the spheres located at atoms and interstitial sites (denoted as X) with majority and minority spin.
	(b) The isosurfaces of the spin charge density with an isovalue of 2.5e-3 $e$/Bohr$^{3}$. 
	Ca, Ga and N atoms are presented as blue, green and pink spheres respectively, while radius of Ca spheres is relatively small in order to clearly show the contributions on charge density. 
	(c) Interstitial charge density at the center of Ca-octahedra with an isovalue is 2.5e-3 $e$/Bohr$^{3}$.  
	(d) and (e) show the electron localization functional maps at (1$\bar{1}$0) and (001) planes, respectively. 
	}
\end{figure}

The orthorhombic phase of Ca$_5$Ga$_2$N$_4$ (space group: $Cmca$, Table \ref{table1}) can be considered as the stacking of [GaN$_2$]$_x$ chain and [Ca$_6$] octahedra \cite{Ca5Ga2N4-1988}. Along [100] direction, each [Ca$_6$] octahedra shares the corner with its neighbors thus forming an infinite chain (Fig. \ref{fig4}). In this structure, half of the [Ca$_6$] octahedra have N filled in the center, while the other half are empty. Therefore, the empty [Ca$_6$] octahedra (Fig. \ref{fig1}c) can accommodate the extra electrons. 

Ca$_5$Ga$_2$N$_4$ is unique because of the unexpected magnetism. We carefully examined possibility of magnetization for all investigated systems with spin polarized calculation. Among these materials, Ca$_5$Ga$_2$N$_4$ was identified as a ferromagnetic phase. The total energy decreases by 65 meV per primitive unit cell (Ca$_5$Ga$_2$N$_4$) by spin polarization due to ferromagnetic (FM) ordering, and the total magnetization moment is 1.25 $\mu$B. We also checked the possibility of various antiferromagnetic (AFM) configurations, the corresponding energy drops are about 60 meV per primitive unit cell, which is slightly less favorable compared to the FM state. Fig. \ref{fig4} shows the spin density of states, the charge density and the ELF maps in order to show its ferromagnetic origin. Remarkably, the magnetization is not only contributed by Ca (4b: 0.5, 0.0, 0.0) and N atoms, but also about 70\% of it from the interstitial sites (4a: 0.0, 0.0, 0.0) (denoted as $X$). Ferromagetism was also reported in some other electride systems, such as $d,f$-metal carbides and nitrides \cite{Y2C-PRB-2015, Inoshita-PRX-2014} in which the partially filled 3$d$ and 4$f$ electronic shells are present. However, the results on light $s$-metal Ca is more interesting. The pronounced ferromagnetic ordering can be explained by the strongly confined electrons, which are localized at the center of Ca-octahedron, as shown in Fig. \ref{fig4}c-e. A close analogous material is the simple cubic phase of elemental potassium at 18.5-22 GPa \cite{Pickard-PRL-2011}. But this prediction has not yet been experimentally confirmed yet. Here, our results suggest that the same phenomenon would be materialized on calcium at ambient condition, with chemical pressure (instead of mechanical pressure) being applied.

\begin{figure*}
\epsfig{file=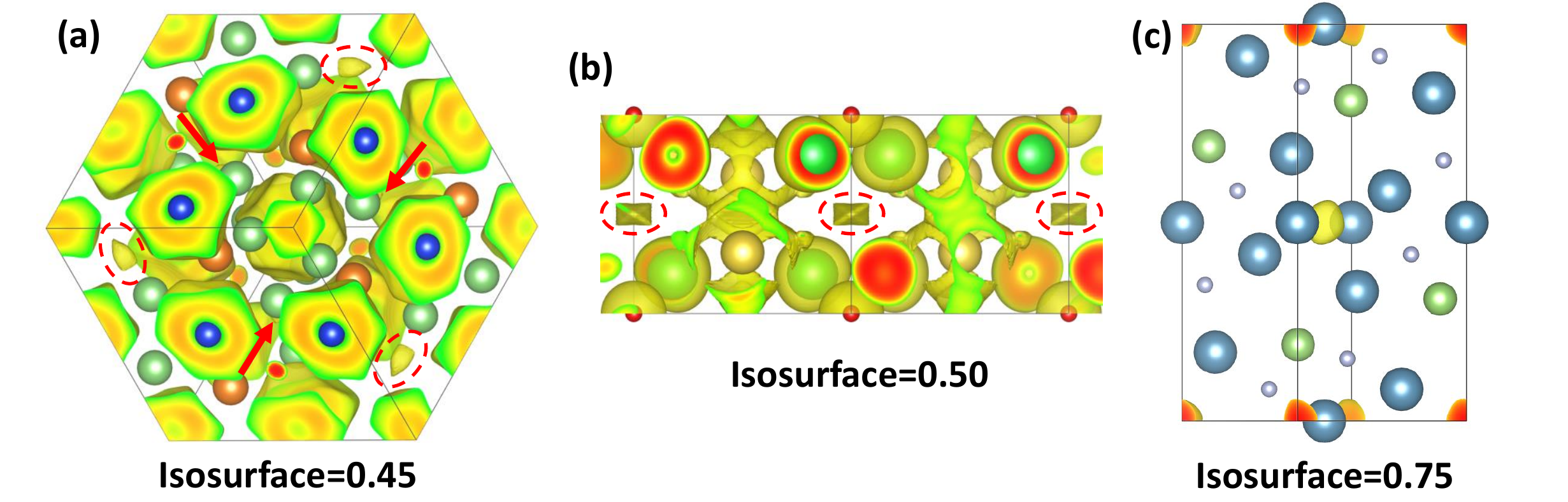, width=0.75\textwidth}
	\caption{\label{fig-all-ELF}
	The calculated electron localization function for (a) Li$_{12}$Mg$_3$Si$_4$:$2e^{-}$, (b) NaBa$_2$O and (c) Ca$_5$Ga$_2$N$_4$ with different isosurface values. 
	The locations of ELF maxima in (a) and (b) are indicated using dashed circles and arrows.} 
\end{figure*}

\subsection{Metrics of the interstitial electrons}
\label{analysis}

Since the prominent feature of electride lies in the high density of floating electrons around the Fermi level, it is instructive to understand how they are confined in the crystal voids for a given electride. Two metrics, i.e., ELF and Bader charge partition, were commonly employed to study the confinement behavior.

ELF \cite{ELF} has been widely used as a powerful tool for analyzing electrides \cite{Li-ANIE-2004}. It was proposed that ELF values larger than 0.75 described well for most of the known electrides, given the fact that most of the known electrides (such as Ca$_2$N and high pressure Na) have strong electron localization \cite{Zhang-PRX-2017}. The ELF of Li$_{12}$Mg$_3$Si$_4$:$2e^{-}$, NaBa$_2$O and Ca$_5$Ga$_2$N$_4$ were also calculated and shown in Fig. \ref{fig-all-ELF}. The electron accumulation at the interstitial sites can be identified from the ELF results. However, the ELF maxima of the corresponding localization attractor of interstitial regions for all systems shows variety of distribution; 0.64 (Ba$_2$NaO), 0.79 (Ca$_5$Ga$_2$N$_4$) and 0.49 (Li$_{12}$Mg$_3$Si$_4$:$2e^{-}$). For reference, C12A7:$e^{-}$ was reported to have an even smaller value ($\sim$ 0.45) \cite{Li-ANIE-2004}. This suggests that strong localization is not a necessary descriptor for electrides. Nevertheless, if the electride does have strong electron localization, the magnetic behavior may appear. The large ELF maxmimum of $X$ sites in Ca$_5$Ga$_2$N$_4$ is a indicator of the ferromagnetism. Similarly, the recently predicted ferromagnetic phase of elemental potassium at 18.5-22 GPa is also due to strong localization\cite{Pickard-PRL-2011}.

Bader analysis has been commonly used in the analysis of interstitial electrons for various electrides \cite{dale-2018, Miao-JACS-2015, Li-Review-2016, Wang-JACS-2017}. We also computed the Bader charges in order to obtain the concentration of floating electrons around the Fermi level. From the band structure and PDOS analysis, it is clear that most of the interstitial electrons are distributed at the range of -1.0 $< E-E_{\textrm F} <$ 0 eV. We thus calculated partial electrons density within the range of -1.0 $< E-E_{\textrm F} <$ 0 eV and partitioned them into different atoms and crystal voids according to Bader scheme\cite{Bader-1990}, in which the zero-flux surface of charge density is used to divide molecular space into atomic volumes \cite{Henkelman-CMS-2006, gudmundsdottir2012local}. The total Bader charges are 0.48 $e$ for Li$_{12}$Mg$_3$Si$_4$:$2e^{-}$ and 0.26 $e$ for C12A7:$e^{-}$. Li$_{12}$Mg$_3$Si$_4$:$2e^{-}$ would have much higher floating electron density of C12A7:$e^{-}$ (about 2 times). Hence Li$_{12}$Mg$_3$Si$_4$:$2e^{-}$  would be desirable for applications if it has comparable stability as C12A7:$2e^{-}$. For Ca$_5$Ga$_2$N$_4$ and Ba$_2$O, the values are 0.44$e$ and 0.55$e$, which are much lower than the formal charges (3$e$ for NaBa$_2$O, and 4$e$ for Ca$_5$Ga$_2$N$_4$). This is understandable since that there also exist plenty of other metallic interactions (such as Ba-Na, Ga-Ga, etc) in those crystals. Although the metallic interaction complicates the scenario, it is evident that the major electron densities close to Fermi level are contributed by the interstitial electrons for all materials investigated in this work, which distinguishes them from the conventional intermetallics.

\begin{figure*}
\epsfig{file=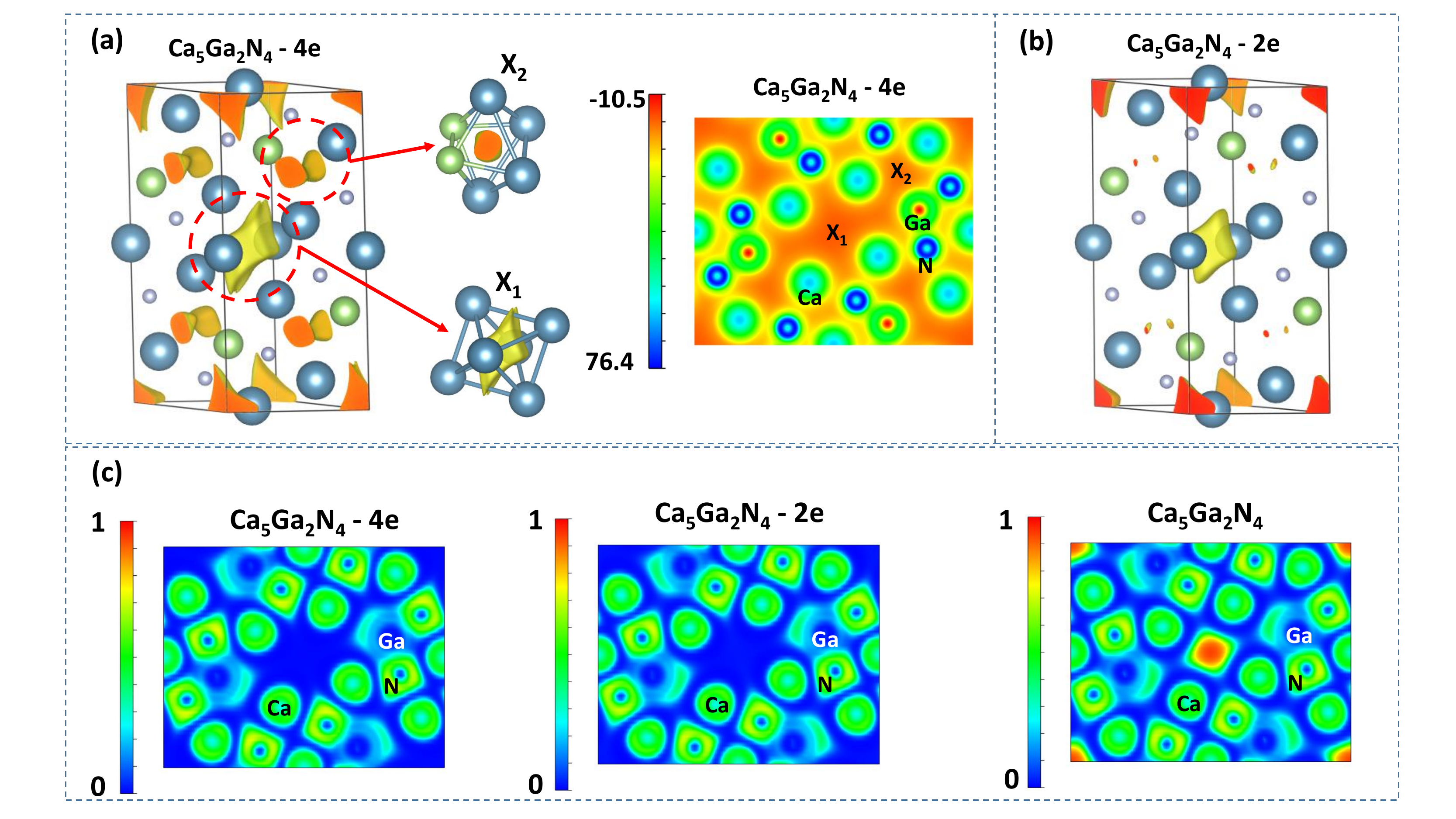, width=0.80\textwidth}
	\caption{\label{fig7}
	The calculated electrostatic potentials for Ca$_5$Ga$_2$N$_4$ after the removal of four (a) and two (b) extra electrons and (c) electron localization functional maps at (110) planes for Ca$_5$Ga$_2$N$_4$ with different charge condition. The isosurface values in (a) and (b) are -0.1 eV.} 
\end{figure*}

\subsection{Competition between metallic bonding and interstitial electrons}
The electronic structure calculations provide us a very interesting picture that the metallic bonding of NaBa$_2$O and Ca$_5$Ga$_2$N$_4$ would host ``extra" electrons by competing with the formation of anionic electrons, whereas only anionic electrons are present in Li$_{12}$Mg$_3$Si$_4$:$2e^{-}$. To get a better understanding of this phenomena, the electrostatic potentials (ESP) of Ca$_5$Ga$_2$N$_4$ after the removal of four and two extra electrons were respectively calculated and shown in Figs. \ref{fig7}a and \ref{fig7}b. Upon the removal of four extra electrons, the lattice of Ca$_5$Ga$_2$N$_4$ shows two kinds of octahedral voids acting as ESP minimum, which are indicated as X$_1$ (surrounded by six Ca atoms) and X$_2$ (surrounded by four Ca and two Ga atoms) in Fig. \ref{fig7}a. The ELF maps of Ca$_5$Ga$_2$N$_4$ with different charge states are plotted in Fig. \ref{fig7}c. It shows that the anionic electrons at X$_1$ sites got disappeared after removing four extra electrons. And the removal of two extra electrons (middle panel of Fig. \ref{fig7}c) shows almost same ELF distribution except the metallic bonding in a crescent shape near Ga atoms is enhanced. This reveals that X$_2$ voids have stronger ability to attract electrons than X$_1$ even though both kinds of voids possess similar ESP values as shown in Fig. \ref{fig7}a. We can understand this phenomena from the fact that X$_2$ voids are surrounded by different types of atoms Ca and Ga. Therefore, the electrons added at X$_2$ will be captured by the Ga atoms with higher electronegativities to form stable metallic bonding. Fig. \ref{fig7}b shows that the ESP minimum at X$_2$ sites become nearly negligible compared with those at X$_1$ sites after adding two electrons. This means that X$_2$ sites are kind of saturated with electrons. Hence, strong electron localization appear at X$_1$ sites after adding two more extra electrons as shown in the right panel of Fig. \ref{fig7}c. The newly added electrons are surrounded by same type of atoms (Ca) and can be stabilized in the voids X$_1$. Therefore, we name the octahedra X$_1$ surrounded by same type of atoms as stable voids (Fig. \ref{fig7}a), whereas X$_2$ are unstable voids.

\begin{figure}
\epsfig{file=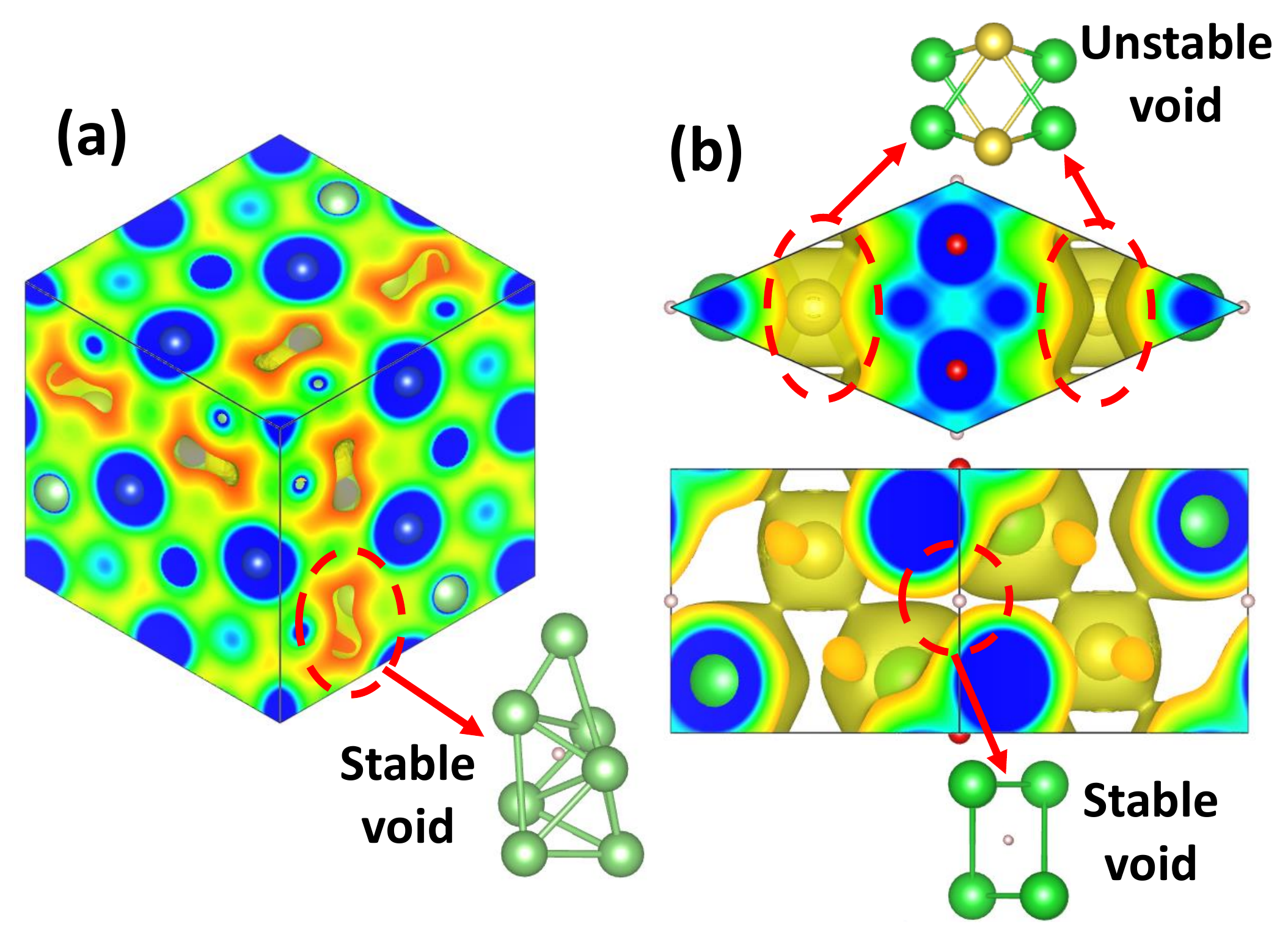, width=0.45\textwidth}
	\caption{\label{fig8}
	The calculated isosurfaces of electrostatic potential for (a) Li$_{12}$Mg$_3$Si$_4$:$2e^{-}$ and (b) NaBa$_2$O after the removal of extra electrons. The isosurface values for the electrostatic potentials are -0.1 eV in both plots. 
		}
\end{figure}

We further computed the ESP distribution of Li$_{12}$Mg$_3$Si$_4$:$2e^{-}$ and NaBa$_2$O after the removal of extra electrons (Fig. \ref{fig8}). Clearly, there is only one type of minimum of ESP in the lattice of Li$_{12}$Mg$_3$Si$_4$:$2e^{-}$, i.e., the interstitial sites surrounded by the Li atoms (Fig. \ref{fig8}a). This means that the extra electrons in Li$_{12}$Mg$_3$Si$_4$:$2e^{-}$ would be pushed to occupy the interstitial sites due to electrostatics. On the other hand, our calculations show that there exist both unstable and stable voids in NaBa$_2$O, suggested by the presence of two types of ESP minima located at the voids surrounded by all Ba atoms (stable voids) and by Na and Ba atoms (unstable voids) (Fig. \ref{fig8}b). Previous study revealed that high external pressure can change the ESP distribution, therefore, can alter the arrangements of electrons in the structure \cite{Rousseau-PRL-2008}. Similar to mechanical pressure, chemical pressure can alter the ESP distribution as well by varying the concentration of electrons. In the present work, our calculations based on three prototypes of electrides reveal that only the stable voids, which are surrounded by same type of atoms, can host anionic electrons. For the co-existence of stable and unstable voids, enough chemical pressure is needed to push the extra electrons to form anionic electrons. Therefore, the number of electrides in ternary compounds is much less than what we expected due to the competition between the metallic bonding at unstable voids and the formation of interstitial electrons at stable voids.

\section{Conclusions}
In sum, we identified three new thermodynamically stable materials (Li$_{12}$Mg$_3$Si$_4$:$2e^{-}$, Ba$_2$NaO and Ca$_5$Ga$_2$N$_4$) as the potential electrides by the rational strategy of high throughput screening based on first-principles calculation. Li$_{12}$Mg$_3$Si$_4$:$2e^{-}$ has similar crystal packing to C12A7:$e^{-}$, but contains 2.5 times higher floating electron density. Ba$_2$NaO is a 1D electride with mixed ionic and metallic bonding, while the 0D electride Ca$_5$Ga$_2$N$_4$ is ferromagnetic due to the strongly localized electrons at interstitials. Since all these compounds have been synthesized in the past, we hope our predictions could stimulate the further experimental validation. These materials represent a class of electrides with excess electrons playing mixed roles of metallic bonding and anionic interstitial electrons. Our discovery bridge the gap between electrides found at ambient and high pressure conditions. Their crystal structures may serve as the new prototypes to design electride materials with different chemical substances. Given the versatile chemistry exhibited in the investigated systems, we believe that there might exist plenty of new electrides for the chemical space which has not been explored in this study. Finally, it is important to note that we restricted the search to only the thermodynamically stable compounds with positive formal charges. However, low energy metastable structures (for instance, diamond) may also be important and survive for a long period. Indeed, some metastable compounds like Li$_{12}$Al$_3$Si$_4$:$2e^{-}$ sharing the same prototype with Li$_{12}$Mg$_3$Si$_4$:$2e^{-}$ was also synthesized in the past \cite{Pavlyuk-1992}.
Therefore, a more complete screening extending to a broader chemical space and stability range is desirable in future.

\begin{acknowledgements}
This work is supported by the National Nuclear Security Administration under the Stewardship Science Academic Alliances program through DOE Cooperative Agreement DENA0001982 and by the Ministry of Education, Culture, Sports, Science and Technology (MEXT) of Japan through the Element Strategy Initiative to Form Core Research Center. JW acknowledges financial support from the Natural Science Foundation of China (Grant No. 51872242). HH acknowledges financial support from MEXT through a Grant-in-Aid for Scientific Research (Grant No. 17H06153). Calculations were performed at the supercomputers of National Institute of Materials Science and Center for Functional Nanomaterials, Brookhaven National Laboratory, which is supported by the U.S. Department of Energy, Office of Basic Energy Sciences, under contract No. DE-AC02-98CH10086. QZ thanks Prof. Ong for providing the code to perform prototype analysis and Prof. Kresse for guides in ESP calculation.
\end{acknowledgements}

\bibliography{reference}

\newpage

\onecolumngrid
\appendix
\begin{table} 
	\caption{The summary of 59 candidate structures considered by our screening after applying the following constraints: 
	1) thermodynamical stability\footnote{We also consider the structures are marginally stable (i.e., less than 1e-3 eV/atom above the convex hull), since this is within numerical accuracy of density functional theory calculation.}
	2) positive formal charge \footnote{Some metals have multiple oxidation states (for instance In has I, II, II, Sn has II and IV), here we excluded the compositions which give neutral charge with different trials of valence states.}
	3) number of atoms in the primitive unit cell is less than 120. 
	The detailed crystallographic information could be found at \href{http://materialsproject.org by material id}
	{http://materialsproject.org} by material id.}
\begin{tabular}{llll}
\hline
\hline
	Material id~~~~~~~~~~~ & System ~~~~~~~~~~~~~~~~~~~   & Space group~~~~~~~~~~~& Prototype~~~~~~  \\
\hline
mp-8331     & Li24Mg6Si8    & I-43d         & Li12Al3Si4   \\
mp-29344    & Li8Ga8Cl24    & Pnma          & LiGaCl3      \\
mp-28327    & Li4Ga4Br12    & P2\_1/m       & LiGaBr3      \\
mp-29345    & Li4Ga4I12     & P2\_1/m       & LiGaBr3      \\
mp-1019519  & Ba4Na2O2      & Cmme          & Ba2NaO       \\
mp-8868     & Ba6Na2N2      & P6\_3/mmc     & BaVS3        \\
mp-28413    & Na8Ga12 Sb12  & Pnma          & Na2(GaSb)3   \\
mp-9378     & Na1Sn2As2     & R-3m          & Ni(AgO)2     \\
mp-28728    & K24Sn12As20   & Pmmn          & K6Sn3As5     \\
mp-28769    & K2Sn4Se8      & Cm            & K(SnSe2)2    \\
mp-646920   & K20As12Pb12   & Pnma          & K5(AsPb)3    \\
mp-505750   & Rb8Sn16Se32   & P2\_1/c       & Rb(SnSe2)2   \\
mp-866492   & Cs18 Al2 O8   & I4/mcm	    & Cs9InO4      \\
mp-582182   & Cs40In24As32  & P2\_1/c       & Cs5In3As4    \\
mp-672338   & Cs4Bi16Te24   & P1            & Cs(Bi2Te3)2  \\
mp-4262     & Be1Al1B1      & F-43m         & TiCoSb       \\
mp-721592   & Ca12Al14O32   & I-43d         & Ca6Al7O16    \\
mp-567191   & Ca5Al5Si5     & P3m1          & CaAlSi       \\
mp-10670    & Ca3Al2Ge2     & Immm          & Ba3(AlGe)2   \\
mp-7263     & Ca2Ga2N2      & P4/nmm        & NbCrN        \\
mp-28489    & Ca10Ga4N8     & Cmce          & Ca5(GaN2)2   \\
mp-541310   & Ca4Ga4Ge4     & P6\_3/mmc     & CaGaGe       \\
mp-510086   & Ca4In2N2      & Cmcm          & Ca2AuN       \\
mp-642331   & Ca19In8 N7    & P1            & Ca19Ag8N7    \\
mp-20419    & Ca8In4N2      & I4\_1/amd     & Ca4In2N      \\
mp-3698     & Sr1Al1Si1     & P-6m2         & SrAlSi       \\
mp-7068     & Sr3Al2Si2     & Immm          & Ba3(AlGe)2   \\
mp-571416   & Sr28Al16Ge6   & R-3           & Sr14Al8Ge3   \\
mp-10671    & Sr3Al2Ge2     & Immm          & Ba3(AlGe)2   \\
mp-13311    & Sr1Al1Ge1     & P-6m2         & SrAlSi       \\
mp-18290    & Sr12Ga10N2    & R-3c          & Ba6Ga5N      \\
mp-972120   & Sr4Ga8As8     & P2/m          & Sr(GaAs)2    \\
mp-641775   & Sr19In8N7     & Fm-3m         & Ca19Ag8N7    \\
mp-19915    & Sr8In4N2      & I4\_1/amd     & Ca4In2N      \\
mp-8539     & Sr2Sn2P2      & P4/nmm        & NbCrN        \\
mp-571291   & Sr36Sn24P48   & Cmce          & Sr3(SnAs2)2  \\
mp-866805   & Sr4Sn12Sb16   & Pnma          & EuSn3Sb4     \\
mp-33444    & Sr5Bi3O12     & P-1           & unknown      \\
mp-9578     & Ba3Al2Si2     & Immm          & Ba3(AlGe)2   \\
mp-13149    & Ba1Al1Si1     & P-6m2         & SrAlSi       \\
mp-10669    & Ba3Al2Ge2     & Immm          & Ba3(AlGe)2   \\
mp-13272    & Ba1Al1Ge1     & P-6m2         & SrAlSi       \\
mp-16859    & Ba12Ga10N2    & R-3c          & Ba6Ga5N      \\
mp-29938    & Ba8Ga16Sb16   & Pnma          & Ba(GaSb)2    \\
mp-605809   & Ba22In12O6    & I4/mcm        & Ba11In6O3    \\
mp-672306   & Ba12In10N2    & R-3c          & Ba6Ga5N      \\
mp-642642   & Ba8In20P20    & Pnma          & Ba2(InAs)5   \\
mp-644875   & Ba8In20As20   & Pnma          & Ba2(InAs)5   \\
mp-601867   & Ba12Sn8P16    & P2\_1/c       & Ba3(SnP2)2   \\
mp-17470    & Ba12Sn8As16   & P2\_1/c       & Ba3(SnP2)2   \\
mp-867542   & Ba16Bi12O42   & P-1           & unknown      \\
mp-758120   & Ba13Bi11O36   & P-1           & unknown      \\
mp-541111   & Al4Ga4Cl16    & P2\_1/c       & BaZnCl4      \\
mp-530331   & Al36Sn6Te60   & P3\_121       & unknown      \\
mp-646102   & Al6Bi10Cl24   & R-3c          & Al3Bi5Cl12   \\
mp-531948   & Ga36Sn6Te60   & P3\_121       & unknown      \\
mp-29075    & Ga6Bi10Cl24   & R-3c          & Al3Bi5Cl12   \\
mp-569237   & Ga12Bi48Cl48  & P6\_3         & Al(BiCl)4    \\
mp-867998   & In15Sn1O24    & R-3           & unknown      \\
\hline
\hline

\end{tabular}
\end{table}

\end{document}